\documentclass[aps,prl,showpacs,showkeys,twocolumn]{revtex4}
\usepackage{graphicx}
\usepackage{amsmath, amsthm, amssymb}

\newcommand{\vect}[1] {\mathbf{#1}}
\newcommand{\dif} {\mathrm{d}}

\newtheorem*{thm1}{Theorem}

\begin{document}

\title{Short Range Scaling Laws of Quantum Gases With Contact Interactions}
\date{\today}
\author{Shina~Tan}
\affiliation{James Franck Institute and Department of Physics,
  University of Chicago, Chicago, Illinois 60637}

\begin{abstract}
The probability amplitude for $N$ particles in a quantum gas with negligible range of interparticle
interaction potentials to come to a small region of size $r$ scales like $r^\gamma$.
It is shown that $\gamma$ is quantitatively related to the ground state energy
of these $N$ fermions in the unitarity limit, confined by an isotropic harmonic potential.
For large $N$, the short range density distribution
of these $N$ particles is predominantly the same as the Thomas-Fermi profile of the gas
in the unitarity limit confined by such a harmonic potential. These results may shed light
on strongly interacting ultracold atomic Fermi gases, in a trap or on an optical lattice.
\end{abstract}

\pacs{03.75.Ss, 05.30.Fk, 05.30.Jp, 71.10.Ca}
\keywords{unitarity limit, scaling law, many-body wave function, quantum gas, contact interaction}

\maketitle


The recent realization of degenerate atomic Fermi gases with large scattering lengths,
near magnetic-field modulated Feshbach resonances
\cite{background1,background2,background3,background4,background5},
 provides motivation to study
a simple model of quantum gases, in which the interaction potentials are of zero range,
and the scattering amplitude between two interacting
particles is characterized by a single parameter, the
s-wave scattering length $a$ \cite{_Landau_book}.
In reality, if the range of interaction potential between
particles is much shorter than the other length scales, the scattering length $a$ and the
typical de Broglie wave length of particles, such a simple model should be justifiable.

An accurate
theory about this model must respect all the features of the many-body wave functions
of this system. 
So the structure of such wave functions deserves in-depth investigations.

In this paper we study an aspect of the short-range structure of these wave functions.
How does the full many-body wave function $\psi$ (regardless of ground state
or any excited state which is present in the density operator at finite temperatures)
scale with $r$, when $N$ of the particles come to a small spatial region of size
 $r\rightarrow 0$? Here $r$ is still much larger than the actual
range of the interaction potential. We expect a scaling law \cite{_Petrov_did_it}
\begin{equation}\label{eq:scaling_definition}
\lim_{r\rightarrow 0}\frac{r}{\psi}\frac{\partial\psi}{\partial r}=\gamma,
\end{equation}
or, more intuitively (but less rigorously), $\psi\sim r^\gamma$. The partial
differential in Eq.~\eqref{eq:scaling_definition} is defined as follows:
1) reduce the distance between each of the $N$ particles and a fixed reference point
$O$ by an amount proportional
to such a distance, without changing the particle's direction with respect to $O$;
2) hold all the other particles at their locations. $r$ is any length scale characterizing
the spatial extension of these $N$ particles.

A major new result in this paper is that $\gamma$ has
an interesting asymptotic formula at large $N$, and it is deeply related to the
ground state energy of the same quantum gas in the unitarity limit \cite{_Ho},
in which $\lvert a\rvert/l$ goes to infinity. $l$ is the average distance between particles.
A broader new finding is that for arbitrary $N$, $\gamma$ is quantitatively related to the
ground state energy of these $N$ particles in the unitarity limit, confined by an
isotropic harmonic potential.

$\gamma$ is a function of $N$, and usually it does not depend on the relative
configuration of these $N$ particles. $\gamma$ does not depend on the scattering
length $a$, because when two scattering particles come to a distance $r\ll \lvert a\rvert$,
the wave function is dominated by the term proportional to $1/r$. $\gamma$ does not depend
on the temperature of the system, either,
because for any energy eigenstate, the typical de Broglie
wave length of a particle is very long compared to the small value of $r$.

In the case of a two-species Fermi gas, with only an interspecies
s-wave contact interaction and a finite scattering length,
$\gamma(1,1)=-1$ trivially \cite{_exception_gamma11}.
Here $\gamma(N_1,N_2)=\gamma(N_2,N_1)$ stands for
the value of $\gamma$ for $N_1$ fermions of one species and $N_2$ ones of the
other species coming to a small region of space. 

$\gamma(1,2)$ was calculated in \cite{Petrov2004PRL},
and its value is $-0.2272757\cdots$. This is associtated with a p-wave
scattering of three fermions, one of species A, and the other two of species B
\cite{Petrov2004PRL}.
If the total orbital angular momentum of these three fermions is precisely
zero, $\gamma=0.1662\cdots$ \cite{Petrov2004PRL}. For a general wave function $\psi$,
this greater value of $\gamma$ only appears at a subset of the configuration space,
with measure zero \cite{_exception_gamma12}.
From this point on, we will only consider the least
possible $\gamma$ in any case.

It is feasible, but nontrivial, to calculate $\gamma(N_1,N_2)$
for all other values of $N_1$ and $N_2$.

Here we study the asymptotic behavior of $\gamma(N/2,N/2)$ at large $N$. We expect,
before we do any concrete calculations, that this asymptotic behavior must be
closely related to the ground state of the 2-component Fermi gas in the unitarity limit.
The reasons are: 1) when $N$ fermions come to a small region in space, $a$ can be regarded
as infinity, also the total energy of these fermions is negligible, compared to the
energy scale set by Heisenberg's uncertainty principle $\hbar^2/mr^2$, where $m$
is the mass of each particle and $\hbar$ Plank's constant divided by $2\pi$,
2) for large $N$, this ``few''-body system should approach the many-body system,
and 3) the least possible $\gamma$ corresponds to the ground state, as will be clear
shortly.

For a little more generality, we consider a $g$-component Fermi gas of particles of identical
 mass $m$, with either contact
interactions or no interactions at all, and approximately calculate
the value of $\gamma$ for $N$ fermions, with $N/g$ of each species, coming to
a small region of space. In a hyperspherical coordinates formalism \cite{_Knirk},
these fermions can be described by a hyperradius $R$, and a set of hyperangles $\Omega$.
\begin{equation}\label{eq:hyperradius}
R^2=\sum_{i=1}^{N}r_i^2,
\end{equation}
where $r_i$ is the distance between the $i$-th fermion and the center-of-mass of the $N$
fermions.
The Schr\"{o}dinger equation for these $N$ fermions is
\begin{equation}\label{eq:Schrodinger}
-\frac{\hbar^2}{2m}\left(\frac{\partial^2}{\partial R^2}+\frac{3N}{R}\frac{\partial}{\partial R}
\right)\psi+\hat{K}\psi=0,
\end{equation}
where the operator $\hat{K}$ is the hyperangular contribution to the total energy. We have
omitted the mass-reduction effect because of the large $N$, and omitted the small constant
term $-4$ in the numerator of the fraction $3N/R$. We also omit
the total energy of the $N$ fermions at small $R$, so the hyperradial part
in the Schr\"{o}dinger equation cancels the hyperangular part.

We need to estimate $K\equiv(\hat{K}\psi)/\psi$, 
in order to estimate $\gamma$. Physically, $K$ should be the lowest
possible total energy of the $N$ fermions at a given $R$, because higher energies lead
to greater values of $\gamma$ which will be overwhelmed by the least $\gamma$
at small distances. So we should use the formula for the ground state energy of the Fermi gas
in the unitarity limit. In the local density approximaion \cite{_Thomas-Fermi},
which should be valid when $N$
is large, the ground state energy per particle is $\dif K/\dif N=(3/5)(1+\beta)
\hbar^2k_F^2/2m$, where $k_F=(6\pi^2\rho/g)^{1/3}$ is the Fermi wave vector, $\rho$
is the local number density of fermions, and
$\beta$ a universal constant \cite{Thomas2002Science},
dependent only of the composition
of the Fermi gas in the unitarity limit. Obviously the density profile of the $N$ fermions
should be isotropic, if we are to achieve the minimum
energy at a given $R$. So $\rho=\rho(r)$, where $r$ is the distance to the center-of-mass.

We now have three equations, for the number of scattering fermions $N$, the hyperradius
$R$, and the energy $K$ at this given hyperradius.
\begin{subequations}\label{eq:NRK}
\begin{align}
N&=\int_0^\infty\rho(r)4\pi r^2\dif r,\\
R^2&=\int_0^\infty r^2\rho(r) 4\pi r^2\dif r,\\
K&=\int_0^\infty\frac{3}{5}\frac{(1+\beta)\hbar^2}{2m}\left(\frac{6\pi^2}{g}\right)^{2/3}
\rho(r)^{5/3}4\pi r^2\dif r.
\end{align}
\end{subequations}

To minimize $K$ at given values of $N$ and $R^2$, we use two
Lagrange multipliers, $\mu$ and $\nu$, and minimize $K-\mu N-\nu R^2$ by adjusting
$\rho(r)$. The only physically meaningful solution to this problem is
\begin{equation}
\rho(r)=
\frac{g}{6\pi^2\hbar^3}\left(\frac{2m\mu}{1+\beta}\right)^{3/2}\Bigl(1-r^2/r_0^2\Bigr)^{3/2}
\end{equation}
for $r<r_0$, and $\rho(r)=0$ for $r>r_0$. Here $r_0=\sqrt{\mu/(-\nu)}$ and $\nu<0$.
The cloud of fermions at the intermediate
stage of scattering (with small instantaneous hyperradius $R$) thus have a density distribution
equivalent to the Thomas-Fermi profile of a Fermi gas in the unitarity limit
confined by an isotropic harmonic potential. 

Expressing $\mu$ and $r_0$ in terms of $N$ and $R^2$, and subsituting the
results into the equation for $K$, we get
\begin{equation}\label{eq:K}
K=g^{-2/3}(1+\beta)(6N)^{8/3}\hbar^2/(128mR^2).
\end{equation}
Note that $K$ is inversely proportional to the hyperradius squared, as a consequence
of the unitarity limit \cite{Thomas2002Science}.
Assuming $\psi\propto R^\gamma$ \cite{_degeneracy},
we then derive from Eq.~\eqref{eq:Schrodinger} that
\begin{equation*}
\gamma(\gamma-1+3N)=g^{-2/3}(1+\beta)(6N)^{8/3}/64.
\end{equation*}
Pauli blocking suppresses the amplitude for many identical fermions to come
to a small region of space, so only the positive solution should be retained, and
\begin{equation}\label{eq:gamma}
\gamma= g^{-1/3}\sqrt{1+\beta}\hspace{.7mm}(6N)^{4/3}/8-3N/2,
\end{equation}
with a relative error likely to be of the order $N^{-2/3}$ at large $N$.

In the case of the 2-component Fermi gas with interspecies contact interaction,
$1+\beta\lessapprox0.44$ according to the latest numerical calculations \cite{Carlson2003PRL}, so
\begin{equation}\label{eq:gamma_2_component}
\gamma(N/2,N/2)\lessapprox0.72 N^{4/3}-3N/2,~~~~~\text{for large $N$}.
\end{equation}

Equation~\eqref{eq:gamma} also holds for the single-component
Fermi gas with \textit{no} interactions, for which $g=1$ and $\beta=0$.
In this case $\gamma(N)$ can be exactly determined for any $N$, as follows.

For $N=1$, the wave function is in most regions
a nonzero value, so $\gamma(1)=0$. For $N=2$, we may consider a Slater determinant
of two base functions in a small region of space, $1$ and $x$, and the wave function
is of the form $x_1-x_2$, where the subscripts label the fermions. So $\gamma(2)=1$.
Actually there are three independent linear functions of $\vect r$, namely $x$, $y$,
and $z$. We may ``fill in'' one more fermion, and form a Slater determinant of,
say, $1$, $x$, and $y$. This means that $\gamma(3)=2$. And similarly
$\gamma(4)=3$, for which the constant ``orbital'' and the three linear ``orbitals''
are all filled by fermions. Note that this kind of short-range analysis is valid
in virtually all regions of the many-fermion configuration space, except regions
of measure zero. 

For five fermions, the constant orbital and the three linear ones are not sufficient. The
fifth fermion has to be filled into a quadratic orbital, say $x^2$. So the Slater
determinant is at least a fifth order polynomial of the coordinates, and $\gamma(5)=5$.
There are six independent quadratic orbitals, $x^2$, $y^2$, $z^2$, $xy$, $yz$, $zx$, and
we can fill them one by one. Each time a more fermion is filled into these orbitals,
$\gamma$ increases by 2. So $\gamma(6)=7$, $\gamma(7)=9$, \dots, and $\gamma(10)=15$.

\begin{figure}
\includegraphics{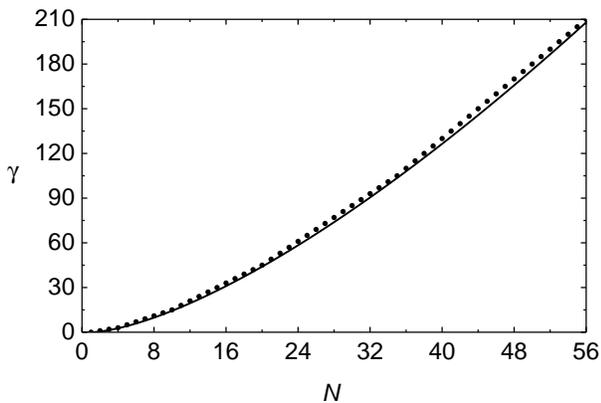}
\caption{The short range scaling exponent $\gamma$ vs. the number of particles $N$, for
a single-component Fermi gas with no interactions. Black dots: the exact $\gamma$ values.
Solid line: the asymptotic formula derived in this paper for large $N$, Eq.~\eqref{eq:gamma}.
\label{fig:gamma}}
\end{figure}

This analysis can be easily generalized to arbitrary $N$. In general, either exactly all
the orbitals of orders no larger than $n$ are filled, \textit{ie},
we have a \textit{short range closed shell structure}, or a portion of the $(n+1)$-th order
orbitals are also filled, \textit{ie}, we have a \textit{short range open shell structure}.
In the former case, $N=(n+1)(n+2)(n+3)/6$ and $\gamma(N)=n(n+1)(n+2)(n+3)/8$.
In the latter case, $N=(n+1)(n+2)(n+3)/6+\delta$, $\gamma(N)=n(n+1)(n+2)(n+3)/8
+(n+1)\delta$, and $1\leqslant\delta<(n+2)(n+3)/2$. In both cases, $\gamma=(6N)^{4/3}/8-3N/2$
with relative errors of the order $N^{-2/3}$ at large $N$. This is completely consistent
with Eq.~\eqref{eq:gamma}. See Fig.~\ref{fig:gamma} for the comparison between the asymptotic
formula and the exact $\gamma$ values.

It is straightforward to verify the validity of Eq.~\eqref{eq:gamma} for the $g$-component
Fermi gas with no interactions.

There is no reason to believe that the favorable comparison between our asymptotic formula
and the actual $\gamma$'s should break down for the two-component Fermi gas with interspecies
long scattering length, although no one has ever computed the actual $\gamma$'s for $N\geqslant 4$.
Here we just show that for $N=2$, Eq.~\eqref{eq:gamma} gives an estimate of $\gamma=-1.19$,
according to the latest $1+\beta$ value \cite{Carlson2003PRL}.
This is already surprisingly close to the actual $\gamma$ value,
$-1$. Note that for a 2-component Fermi gas with no interactions, our asymptotic formula gives
$\gamma(1,1)=-0.27$, while the actual $\gamma(1,1)=0$. The magnitudes of errors are apparently similar
for these two different systems. If we solve Eq.~\eqref{eq:gamma} for $1+\beta$'s from the known exact
$\gamma(1,1)$ values, we get $1+\beta=0.538$ for the 2-component Fermi gas with interspecies long
scattering length, and $1+\beta=1.21$ for the 2-component Fermi gas with no interactions. Both of these
two estimates are larger than the actual $1+\beta$ values by just about 21-22\%, at such a small
number of fermions as $N=2$.

We may generalize Eq.~\eqref{eq:gamma} to the case in which the numbers of fermions coming
to a small spatial region are not equal among different species, but have some fixed ratio.
In this case, we expect to approach the ground state of the Fermi gas in the unitarity limit
with that ratio among the populations of different species, and will get some different $\beta$.
Also we can study the case in which different components have different masses. 

We may also generalize the same analysis to other quantum gases with contact interactions,
as long as those gases are stable. A prominant example is a Fermi-Bose mixture, with
no interactions between the bosons. If the particles all have the same mass,
and the number of colliding particles is evenly distributed among all the species,
the formula for $\gamma$ is formally
the same as Eq.~\eqref{eq:gamma}, except that the value of $\beta$ is different.

Our results are a demonstration of quantum parallelism in the context of quantum gases
with long scattering lengths. Normally one regards the regimes in which $k_F a\sim O(1)$
as distinct from the unitarity limit, in which $k_F a=\infty$. We have shown, however,
that such a simple picture is not complete. In fact, even for finite $k_F a$, small portions of
particles still have finite, albeit very small, probability amplitudes of \textit{clustering}
for short times, and when they do so, their behavior is very similar to the behavior
of the \textit{entire} quantum gas in the unitarity limit. 
We may call this peculiar behavior \textit{instantaneous unitarity limit}.
Note that such a phenomenon coexists with other more familiar ones
in \textit{any} many-body wave functions, regardless of the ground state or excited states.
Non of the many-body theories built to date reflect this feature.

Our short-range structure analysis of the many-body wave function
may shed light on the theory of the quantum gases away from the unitarity limits. One interesting
question is: how to construct an accurate theory of a 2-component Fermi gas in the pair-wise
Bose-Enstein condensation (BEC) states, when $k_F a$ evolves from small positive numbers to values
of order unity? In a low-density expansion in the BEC regime, one is forced to take progressively
higher orders of correlations into account, with more and more dimers colliding
at distances of the order $a$. 
One has to evaluate all these scattering amplitudes, when the density is not
extremely low. While these amplitudes may be easier to calculate directly
when the number of colliding dimers
is not large, at large numbers of dimers, one will find the idea of instantaneous unitarity limit,
or, more generally, instantaneous high density, helpful.

Another possible application of the ideas in this paper is: we may refine our asymptotic
formula, by taking all the correction terms to the Thomas-Fermi approximation
into account; then, by solving a quantum mechanical
problem of a small number of fermions, we could accurately determine the physical properties of the
gas in the unitarity limit. 

In the end of this paper, it should be pointed out that there is an
\textit{exact short range scaling theorem},
which places Eq.~\eqref{eq:gamma} on a broader and more rigorous ground.

\begin{thm1}
The ground state energy of $N$ fermions of identical mass $m$ with contact interactions of
infinite scattering lengths or no interactions,
confined in an isotropic harmonic trap of angular frequency $\omega$, is
\begin{equation}\label{eq:scaling_theorem}
E=(\gamma+3N/2)\hbar\omega~~~~~~~~~~\text{(exact)},
\end{equation}
where $\gamma$ is the short range scaling exponent defined previously.
\end{thm1}
\begin{proof}
Let us start with the wave function of $N$ fermions with contact interactions of finite
scattering lengths or no interactions, $\psi(\vect r_1,\cdots,\vect r_N)$. When $\vect r_i\rightarrow0$
but the distance between any two interacting fermions remains nonzero, $\psi$ satisfies the scaling law
and the Schr\"{o}dinger equation
\begin{subequations}\label{eq:short_distance}
\begin{align}
\sum_{i=1}^{N}\vect r_i\cdot\nabla_i\psi        &=\gamma\psi,\\
-\frac{\hbar^2}{2m}\sum_{i=1}^{N}\nabla_i^2\psi &=0.
\end{align}
\end{subequations}
$\psi$ also must satisfy
appropriate boundary conditions whenever the distance $r$ between two fermions vanishes,
and is like $c/r+O(r)$ for $a/r\rightarrow\infty$, or $c+O(r)$ for $a=0$,
where $c$ is independent of $r$.
We now extrapolate the $N$-fermion short-distance wave function $\psi$
to finite distances according to the scaling law, and then define
\begin{equation}
\psi'\equiv\psi\exp\left(-\frac{m\omega}{2\hbar}\sum_{i=1}^{N}r_i^2\right),
\end{equation}
which clearly has the same symmetry and boundary conditions as $\psi$. It is straightforward to
derive from Eq.~\eqref{eq:short_distance} that for nonvanishing distances
between interacting fermions
\begin{equation}
\sum_{i=1}^{N}\left(-\frac{\hbar^2}{2m}\nabla_i^2+\frac{1}{2}m\omega^2r_i^2\right)
\psi'=(\gamma+3N/2)\hbar\omega\psi'.
\end{equation}
So for the lowest possible $\gamma$, $\psi'$ is the ground state wave function of
the $N$ fermions (with zero or infinite scattering lengths) in the harmonic trap, with the
energy given by Eq.~\eqref{eq:scaling_theorem}.
\end{proof}

At large $N$, we can rederive Eq.~\eqref{eq:gamma} by using Eq.~\eqref{eq:scaling_theorem}
and the Thomas-Fermi approximation \cite{_Thomas-Fermi} of $E$ \cite{_E}.
We see that the term $-3N/2$ in Eq.~\eqref{eq:gamma} is deeply linked to the zero-point
energies of harmonic oscillators.


\begin{thebibliography}{17}
\bibitem[{_ex({\natexlab{a}})}]{_exception_gamma11}
\bibinfo{note}{If the total orbital angular momentum of these two particles
  happens to have zero probability of being zero, $\gamma(1,1)$ is a positive
  integer}.
\bibitem[{_de()}]{_degeneracy}
\bibinfo{note}{If the lowest eigenvalue of $\hat{K}$ has a degree of degeneracy
  $s\geqslant2$ with the same eigenfunction, the wavefunction may behave like
  $R^{\gamma}$ times an $(s-1)$-st order polynomial of $\ln R$. This caveat
  does not affect our estimate of $\gamma$}.
\end{thebibliography}
Financial support by the Department of Physics, University of Chicago is gratefully acknowledged.
The author thanks K. Levin for introducing the area of cold Fermi gases to him, and thanks K. Levin and
C. Chen for comments.

\end{document}